 \documentclass{article}

\usepackage{epstopdf}% To incorporate .eps illustrations using PDFLaTeX, etc.
\usepackage[caption=false]{subfig}% Support for small, `sub' figures and tables

\usepackage[numbers,sort&compress]{natbib}% Citation support using natbib.sty
\bibpunct[, ]{[}{]}{,}{n}{,}{,}% Citation support using natbib.sty
\makeatletter% @ becomes a letter
\def\NAT@def@citea{\def\@citea{\NAT@separator}}% Suppress spaces between citations using natbib.sty
\makeatother% @ becomes a symbol again

\newtheorem{theorem}{Theorem}[section]

\newtheorem{proposition}[theorem]{Proposition}

\usepackage{amsfonts} 
\usepackage{amssymb,amsmath} 
\usepackage{graphicx,color}

\usepackage[breaklinks,colorlinks=true,linkcolor=blue,citecolor=red,backref=page]{hyperref}

\DeclareMathAlphabet{\itbf}{OML}{cmm}{b}{it}

\def\bq{{{\itbf q}}}

\def\br{{{\itbf r}}}

\def\by{{{\itbf y}}}
\def\bx{{{\itbf x}}}

\def\bk{{{\itbf k}}}

 \def\bzeta{{\boldsymbol{\zeta}}}
\def\bxi{{\boldsymbol{\xi}}}

 \def\eps{{\varepsilon}}
\def\epst{{\epsilon}}

\newcommand{\RR}{\mathbb{R}}

\newcommand{\EE}{\mathbb{E}}
\newcommand{\cA}{\mathcal{A}_{\rm tr}}
 
  \newcommand{\ean}{\end{eqnarray*}} 
\newcommand{\ban}{\begin{eqnarray*}} 
 \newcommand{\ea}{\end{eqnarray}} 
\newcommand{\ba}{\begin{eqnarray}}

\begin{document}  
 
 \title{Partially Coherent  Electromagnetic Beam  Propagation in Random Media}

\author{Josselin
Garnier\thanks{\footnotesize Centre de Math\'ematiques Appliqu\'ees, Ecole Polytechnique,
91128 Palaiseau Cedex, France (josselin.garnier@polytechnique.edu)} 
\and Knut S\o lna\thanks{\footnotesize Department of Mathematics, 
University of California, Irvine CA 92697
(ksolna@math.uci.edu)}
}

% \author{
%\name{Josselin Garnier\textsuperscript{a}  and Knut S\o lna\textsuperscript{b}}
%\affil{\textsuperscript{a}Centre de Math\'ematiques Appliqu\'ees, Ecole Polytechnique, 91128 Palaiseau Cedex,
%France
%{\tt josselin.garnier@polytechnique.edu}; \textsuperscript{b}Department of Mathematics, 
%University of California, Irvine CA 92697
%{\tt ksolna@math.uci.edu}}
%}
% 

\maketitle
 
\begin{abstract}
A theory for the characterization of the fourth moment of electromagnetic  wave
beams is presented in the case when the source is partially coherent. 
A Gaussian-Schell model is used for the partially coherent random source. 
The white-noise paraxial regime is considered, which holds when  the wavelength is much smaller 
 than the correlation radius of the source, the beam radius of the source, and the correlation length of the medium, which are themselves much smaller than the propagation distance.
The complex wave amplitude field can then be described by the  
It\^o-Schr\"odinger equation.  
This equation gives closed evolution equations for the
wave field moments at all orders and here the fourth order equations are considered.  
The general  fourth moment equations are solved explicitly
in the scintillation regime 
(when the correlation radius of the source is of the same order as the correlation radius of the medium, but the beam radius is much larger)
and  the result gives a characterization 
of the intensity covariance function.  The form of the intensity covariance function 
derives from the solution of the  transport equation  for the Wigner distribution
associated with the second wave moment. 
The fourth moment results for polarized waves is  used in an application to imaging of partially 
coherent sources. 
 \end{abstract}

%\begin{keywords}
%Random media, electromagnetic waves, multiple scattering,  Schr\"odinger equation, Gauss-Schell Model, scintillation,
%source imaging.  
% \end{keywords}

\newpage

%\begin{AMS}
%60H15, 35R60, 74J20.
%\end{AMS}

\pagestyle{myheadings}
\thispagestyle{plain}
\markboth{Partially Coherent  Electromagnetic Beam  Propagation in Random Media}{J. Garnier and K. S\o lna}

\tableofcontents

\section{Introduction} 

We consider beam wave propagation in  complex media in the situation when we model 
both the source and the medium as being random.  
Modeling with a random or complex medium in the context of wave propagation 
is natural in  many situations. In the early  foundational work \cite{t1,t2} 
and also in \cite{andrews,ishimaru} a main motivation 
was propagation  through the turbulent atmosphere, but there are many other 
important applications as well. In cases where one considers 
 propagation through  the fluctuating ocean,
the earth's crust or through biological tissue it is also natural and convenient 
in many cases to model in terms 
of a random medium \cite{book1,ishimaru,korot}.  
In these cases the medium may be too complex to describe
pointwise, but one can hope to be able to describe or model the statistics of the medium fluctuations. 
The challenge is then to capture the complicated   coupling between the medium fluctuations
and the wave field and to understand how  a particular model for the random medium 
statistics  affects the statistics of the wave field which becomes a random field due to  
random scattering.  
In this paper we consider the case of beam waves or paraxial waves in a high frequency
long range propagation scenario, when  the wavelength is much smaller 
 than the correlation radius of the source, the beam radius of the source, and the correlation length of the medium, which are themselves much smaller than the propagation distance.
 In this case we can approximate the wave field
in terms of  the solution of the It\^o-Schr\"odinger equation.
This equation was analyzed in \cite{dawson} and derived from the Helmholtz equation in \cite{gar09}.  
Despite the long history of the  theory of waves in random media 
a rigorous and explicit description of the fourth moment  was only  obtained
in \cite{gar16a}  in the scintillation regime 
(when the correlation radius of the source is of the same order as the correlation radius of the medium, but the beam radius is much larger).
 Here we generalize this 
fourth moment theory to the case of polarized waves.   
This is a deep and quite surprising result in the theory of waves in random media.  
It states that even though the polarized wave has only  partially lost its coherence
due to scattering   it behaves from the point of view of the fourth moment as if 
it was a Gaussian field.   In some sense this quasi-Gaussian property  explains some of the success
or robustness of the theory of waves in random media since the second-order
characterization,  which in general is relatively easy to obtain,   also explains the behavior of fourth-order wave functionals.     
 
An important aspect of our modeling in this paper is that we also consider
the source as being random.  
Modeling with a random source may be motivated 
by the complexity of the source as when one considers  emission from a star.
A second motivation for  understanding  and analyzing  beam wave propagation from a random source
is that such sources have been promoted as being desirable for scintillation reduction 
when beaming through a complex medium, see  \cite{gbur,svetlana2,svetlana3,gbur2,voelz,pcrev}
and references therein. Scintillation here refers to the situation that
the transmitted beam intensity fluctuates rapidly due to scattering over the propagation path.
One  intuition is that by using a complex source one gets  a better mixing over wave ray paths
and a scintillation reduction.  Our objective here is to develop  an analytic framework where
in particular such questions can be analyzed.    We consider  the Gaussian-Schell model for the source when the source coherence  statistics is defined in terms
of Gaussian envelopes. This gives rather simple and  convenient forms for the wave statistics,
but the theory can easily be modified to more general forms for the source 
statistics.     The situation with a partially coherent source, but a homogeneous medium 
was considered in \cite{gar00},  while we here consider the case when also the medium 
is random.  Note that 
  we focus on the fourth moment, while
 in \cite{fps} moments of all orders were considered under some simplifying 
assumptions. The assumption that  allows us to get explicit expressions   for the
fourth moment  is to assume  the scintillation regime,
when the correlation radius of the source is of the same order as the correlation radius of the medium, but the beam radius is much larger.   
Fluctuations of scalar wave intensity 
was considered in \cite{ryzhikDC,gu21} for a beam type propagation when the random medium fluctuations
are Gaussian, while we assume Gaussianity for the source, but not for the medium.  In \cite{bal} a characterization of
intensity fluctuation and how it depends on the regularity 
of deterministic initial data is presented.  
Here we consider the case with random initial data
with smooth Gaussian statistics.

We remark that there are a number of approaches to model
high-order moments of the wave field that are based on perturbative
approaches. Indeed,  the derivation of such approximations is based
on the premise that the waves is only perturbed or affected by the scattering to
lower order \cite{andrews,baker,char1}. In this paper we describe an analytic framework that gives a rigorous
scaling limit identification of the fourth-order moment in the saturated regime when the incoherent or scattered part of the wave field 
is of order one and  is as large as or larger than  the coherent component. 
In fact this description also captures the situation when the wave is fully 
incoherent and the  coherent part of the wave energy is essentially fully scattered.         
  
We describe next a main result in the paper.
We consider  the situation when the time-harmonic electric field in the  source plane  $z=0$
 correspond to a partially coherent beam and has the form 
\begin{equation}
\vec{\itbf E}(z=0,\bx) = 
\sum_{j=1}^2 f_j(\bx) \hat{\itbf e}_j   ,
\end{equation}
where $\hat{\itbf e}_1$ and $\hat{\itbf e}_2$ are two orthogonal unit vectors in the transverse plane,
$z$ is the beam propagation direction and $\bx$ the lateral spatial coordinates. 
 The functions  $f_1$, $f_2$ are zero-mean Gaussian processes with covariance
\begin{equation}
\EE\big[ f_j\big(\br+\frac{\bq}{2}\big)\overline{f_l}\big(\br-\frac{\bq}{2}\big) \big]
=
\left\{
\begin{array}{ll}
A_j^2 \exp\Big( - \frac{|\br|^2}{r_o^2}-\frac{|\bq|^2}{4\rho_o^2}\Big) &\mbox{ if } j=l ,\\
A_j A_l \chi  \exp\Big( - \frac{|\br|^2}{r_o^2}-\frac{|\bq|^2}{4\rho_1^2}\Big) &\mbox{ if } j\neq l ,
\end{array}
\right.
\label{eq:covsource}
\end{equation}
where the polarization degree $\chi\in [-1,1]$,  $r_o$ is the beam radius of the source, $\rho_o$ is the correlation radius of the source and we have $\rho_1\leq \rho_o$.
We refer to \cite{born} for further background on modeling with  polarized waves. 
Consider first the second field moment in the form of the mutual coherence function 
in the white-noise paraxial regime (with   the wavelength  smaller than the correlation radii of the medium and the beam which
are moreover on the scale of the   the beam radius, 
which in turn is smaller than the propagation distance, see Appendix \ref{app:a}).
As discussed in Section \ref{sec:7}  
the mutual coherence function of the wave field is  in this regime
 given by
\begin{equation}
\mu_{2,jl}(z,\br,\bq) 
=
\EE \big[u_j\big(z,\br+\frac{\bq}{2}\big) \overline{u_l}\big(z,\br-\frac{\bq}{2}\big)\big]  =
\left\{
\begin{array}{ll}
A_j^2  {\cal H}_{\rho_o}(z,\br,\bq) &\mbox{ if } j=l ,\\
A_j A_l \chi   {\cal H}_{\rho_1}(z,\br,\bq) &\mbox{ if } j\neq l ,
\end{array}
\right.
\label{eq:covfield}
\end{equation}
where the fundamental second-order lateral scattering function is defined by
\begin{align}
\nonumber
 {\cal H}_{\rho_o}(z,\br,\bq) =&
 \frac{r_o^2}{4\pi} \int_{\RR^2}  \exp\Big( i \bzeta \cdot\br - \frac{r_o^2 |\bzeta|^2}{4} - \frac{|\bzeta \frac{z}{k_o}-\bq|^2}{4\rho_o^2} \Big)\\
& 
\times \exp\Big( \frac{k_o^2}{4} \int_0^z C\big( \frac{\bzeta z'}{k_o} - \bq\big)-C({\bf 0}) dz'\Big) d\bzeta ,
\label{def:Hq}
\end{align}
with $k_o$ the central wavenumber and $C$ the 
covariance function of the medium fluctuations when integrated in the $z$-dimension (see Eq.~(\ref{def:C2})).   
The intensity is defined by 
\begin{equation}
\label{def:int}
I(z,\br) = \sum_{j=1}^2 |u_j(z,\br)|^2 ,
\end{equation}
 and  the mean intensity is then 
\begin{equation}
\EE \big[ I ({z}, {\br}  )  \big]
=(A_1^2+A_2^2 ) {\cal H}_{\rho_o}(z,\br,{\bf 0})   .
\label{eq:expmeanint}
\end{equation}
Consider next the fourth field moment in form of the  covariance of the intensity.
As discussed in Section \ref{sec:9},
in the scintillation regime where the wavelength is smaller than the correlation radii of the medium and the beam, which are smaller than the beam radius, which is smaller than the propagation distance,
 the intensity covariance  has the form 
\ba\nonumber 
 {\rm Cov}\big( I\big({z}, {\br} +\frac{\bq}{2}\big) , I\big({z}, {\br}-\frac{\bq}{2}\big)\big)
&=& 
\big(A_1^4+A_2^4\big) |{\cal H}_{\rho_o}(z,\br,\bq)|^2  \\ && \hbox{}
+
2A_1^2A_2^2  \chi^2 |{\cal H}_{\rho_1}(z,\br,\bq)|^2 .
\label{eq:expcovint}
\ea
% This result and the scaling assumptions leading up to it is detailed in Section  \ref{sec:7} and Appendix \ref{app:A}.  
This representation proves the quasi-Gaussian property, that the fourth moment
(intensity covariance) derives from the second moment (the mutual coherence function), as is the case 
in general for Gaussian random fields.  
From this description  we can identify the intensity decoherence scale
and spreading scale and we discuss this explicitly  in the strong scattering case  in Appendix  \ref{sec:5}.
More generally the quasi-Gaussian property provides the basis for analysis 
of a number of wave propagation challenges and we discuss one such application 
in Section \ref{sec:app}.      We remark that  fourth moments of the  wave fields that are more general
than the intensity covariance can be obtained via
Proposition \ref{prop:3} in Appendix \ref{sec:6} and that the quasi-Gaussian property holds
for such general moments.

The outline of the paper  is as follows. 
In Section \ref{sec:7}  we describe modeling of the 
partially coherent source in the polarized case and we 
present the It\^o-Schr\"odinger equation describing   wave 
propagation in the white-noise paraxial regime.  
In  Section
\ref{sec:9} we present the fourth moment result  for  the
polarized waves.  The main result that allows us to
do this is that the It\^o-Schr\"odinger equations for the polarization modes
are dynamically uncoupled,  however,  statistically coupled.       
   In Section \ref{sec:app} we discuss an application of the theory
that we have developed to source imaging (using the intensity covariance function instead of the mutual coherence function as in \cite{ip}). 
The details of the derivation and also a discussion of the second-order moment
or mutual coherence function can be found in the appendices.

 \section{Electromagnetic Waves in the White-Noise Paraxial Regime}\label{sec:7} 
 
 We consider   propagation of a partially coherent electromagnetic  beam waves 
 through a three-dimensional  random medium.
% The characteristic scaling regime we will consider is  the white-noise paraxial regime (when the wavelength is much smaller 
% than the beam radius of the source, the correlation radius of the source, and the correlation length of the medium, which are themselves much smaller than the propagation distance).
   Maxwell's equations for the three-dimensional electric field $\vec{\itbf E}$ and 
the three-dimensional magnetic field strength  $\vec{\itbf H}$ are in the time-harmonic case (with frequency $\omega_o$):
\begin{eqnarray}
\label{eq:0a}
&&\nabla \times \vec{\itbf E} = - i \omega_o \mu(z,\bx) \vec{\itbf H}, \\
&&\nabla \cdot ( \epst(z,\bx) \vec{\itbf E} ) =\rho(z,\bx), \\
&&\nabla \times \vec{\itbf H} = \vec{\itbf J}^{(s)}(z,\bx) + i \omega_o \epst(z,\bx)  \vec{\itbf E},    \\
&& \nabla \cdot ( \mu(z,\bx) \vec{\itbf H}) =0 .  
\label{eq:0d}
\end{eqnarray}
 The term $\vec{\itbf J}^{(s)}$ is a current source term,
 $\epst$  is the dielectric permittivity of the medium, and
$\mu$ is  the magnetic permeability of the medium.
Note that the equation of continuity of charge
$ i \omega_o  \rho + \nabla \cdot \vec{\itbf J}^{(s)} =0$
is automatically satisfied.

We assume that
\begin{itemize}
\item
 The medium is randomly heterogeneous:
 \begin{align}
\label{eq:med10}
\epst(z,\bx)
  &=
  \epst_o \big[  1+
         m_\epst( z,\bx )   \big]
            , \\
 \mu(z,\bx)&=
    \mu_o\big[    1+
           m_\mu(z,\bx) \big]     .
           \label{eq:med20} 
\end{align}
The  random processes  $m_\epst(z,\bx)$ and $m_\mu(z,\bx)$  are  bounded, stationary, and zero-mean and they 
satisfy ergodic (mixing) conditions in $z$. 

\item 
We consider a partially coherent source $\vec{\itbf f}(\bx)$, which is a field with Gaussian statistics and mean zero that is localized in the plane $z=0$.
We address the case of a Gauss-Schell model for the source.  
The source is then $\vec{\itbf J}^{(s)}(z,\bx) = 
2 \mu_o^{-1/2}\epst_0^{1/2}
\vec{\itbf f}(\bx)  \delta(z)$, where $f_3=0$ and $f_1$, $f_2$ are zero-mean Gaussian processes with covariance
\begin{equation}\label{eq:ic} 
\EE\big[ f_j\big(\br+\frac{\bq}{2}\big)\overline{f_l}\big(\br-\frac{\bq}{2}\big) \big]
=
\left\{
\begin{array}{ll}
A_j^2 \exp\Big( - \frac{|\br|^2}{r_o^2}-\frac{|\bq|^2}{4\rho_o^2}\Big) &\mbox{ if } j=l ,\\
A_j A_l \chi  \exp\Big( - \frac{|\br|^2}{r_o^2}-\frac{|\bq|^2}{4\rho_1^2}\Big) &\mbox{ if } j\neq l .
\end{array}
\right.
\end{equation}
All parameters are real and positive, with $\chi\in [-1,1]$.
The parameters have to satisfy several constraints to ensure that we deal with a well-defined covariance function:
$$
\rho_o \leq r_o,\quad \quad 
\chi^2 \rho_o^{-2}+(1-\chi^2) r_o^{-2} \leq \rho_1^{-2} \leq  \rho_o^{-2} .
$$
\end{itemize}
Here $r_o$ is the radius of the beam and $\rho_o$ is the   correlation radius.
The special case $\rho_o=r_o, \chi=0$
corresponds to a coherent source with components $f_j(\bx) = f_{o,j} \exp(-|\bx|^2/(2r_o^2))$ where $f_{o,j},  j=1,2$ are  two
independent complex-valued  zero-mean Gaussian random variables with variance $A_j^2$. 
In the general case with  $\rho_o < r_o$ 
the field is partially coherent and the  field components have the form of correlated 
 speckle patterns with speckle spots with typical radius
 $\rho_o$ and with an overall intensity envelope that is  $\exp(-|\br|^2 / r_o^2)$.

In the white-noise paraxial regime 
(which holds when the wavelength is much smaller than the correlation radius of the source, the correlation radius of the medium, and the beam radius, which are themselves much smaller than the propagation distance)
the electric field  modulo   a range-dependent phase (see Appendix \ref{app:a}) has the form 
\begin{equation}
\vec{\itbf E}(z,\bx) = 
\sum_{j=1}^2 u_j(z,\bx) \hat{\itbf e}_j  ,
\end{equation}
where $\hat{\itbf e}_1$ and $\hat{\itbf e}_2$ are the unit vectors in the transverse plane pointing in  the $x$ and $y$ directions
and
the complex amplitude fields $u_j$  are
the solution of the following statistically coupled    
  It\^o-Schr\"odinger equations   \cite{gar09,59,85}:
  \begin{equation}
\label{eq:model2}
 d {u}_j(z,\bx)   =     
          \frac{ i }{2k_o} \Delta_{\bx}   {u}_j(z,\bx) dz
   +   \frac{ik_o}{2}   {u}_j (z,\bx ) \circ  d{B}(z,\bx) 
  ,  \quad j=1,2,
\end{equation}
with the initial condition in the plane $z=0$:
$$
 {u}_j(z= 0,\bx )  = f_j(\bx) .
$$
  Here  the random process $B(z,\bx)$ is a real-valued  Brownian field with   
a covariance that derives from two-point statistics in the model for the 
medium fluctuations  in (\ref{eq:med10}-\ref{eq:med20})
 \begin{equation}
 \label{defB2}
\EE[   {B}(z,\bx)  {B}(z',\bx') ] =  
 {\min\{z, z'\}}    {C}(\bx - \bx') ,
\end{equation}
where 
\ba\label{def:C2}
  C(\bx) = \int_\RR  \EE[(m_\epst+m_\mu)(\bx'+  \bx,z'+z)(m_\epst+m_\mu)(\bx',z')] dz . 
 \ea   
 Note therefore that the evolution equations for the lateral components of the
 electromagnetic  field is driven by the same Brownian field. 
 We remark that  $C({\bf 0})$ can be interpreted as the product of the variance of the fluctuations of the 
 random medium times its longitudinal correlation radius:
 $$
 C({\bf 0}) = \sigma^2 \ell_z  ,  
 $$ 
 for $\sigma$ the standard  deviation of the medium fluctuations: $\sigma^2=\EE[ (m_\epst+m_\mu)^2(\bx',z') ]$.
   
 The derivation of  (\ref{eq:model2}) from the random three-dimensional scalar wave equation is presented in \cite{gar09}. Its derivation  for the Maxwell's equations 
 (\ref{eq:0a}-\ref{eq:0d})
 is presented in
 \cite{59,85}. 
In Eq.~(\ref{eq:model2})  the symbol $\circ$ stands for the Stratonovich stochastic integral \cite{gar09}.
The first- and second-order moments of the wave field solution of (\ref{eq:model2}) have been studied in   \cite{gar09,59} 
 and recover results derived in \cite{ishimaru}.
In view of the centered partial coherent source 
we find that the first-order moment of the wave field is zero. 
The governing equations for the higher  order moments   
can  be identified via It\^o calculus for Hilbert space valued processes.
We find in particular that the second-order moment of the wave field (mutual coherence function) defined by
\begin{equation}
\mu_{2,jl}(z,\br,\bq) = \EE\Big[ u_j\big(z,\br+\frac{\bq}{2}\big) \overline{u_l\big(z,\br-\frac{\bq}{2}\big)}\Big]
\end{equation}
satisfies  \cite{gar14a}
\begin{equation}\label{eq:mu2} 
\frac{\partial \mu_{2,jl}}{\partial z} = \frac{i}{k_o}  \nabla_{\br} \cdot\nabla_{\bq}  \mu_{2,jl} + \frac{k_o^2 }{4} {U}_{2} \big(  \bq \big)
 \mu_{2,jl}   , 
\end{equation}
with the potential $U_2(\bq) =  C(\bq)-C({\bf 0})$ and the initial condition 
determined by (\ref{eq:ic}). 
As shown in Appendix~\ref{sec:4}, it then follows that
the mutual coherence function is given by
\begin{equation}
\mu_{2,jl}(z,\br,\bq) 
%= \EE \big[u_j\big(z,\br+\frac{\bq}{2}\big) \overline{u_l}\big(z,\br-\frac{\bq}{2}\big)\big]  
=
\left\{
\begin{array}{ll}
A_j^2  {\cal H}_{\rho_o}(z,\br,\bq) &\mbox{ if } j=l ,\\
A_j A_l \chi   {\cal H}_{\rho_1}(z,\br,\bq) &\mbox{ if } j\neq l ,
\end{array}
\right.
\label{eq:covfield2}
\end{equation}
where ${\cal H}_{\rho_o}(z,\br,\bq) $ is defined by (\ref{def:Hq}).

We next show how our main quantity of interest  - the intensity covariance -
can be expressed in terms of the mutual coherence function. 
     
\section{The Intensity Covariance for Polarized Waves}
\label{sec:9}%
The intensity is defined by (\ref{def:int}) and 
by (\ref{eq:covfield}), its mean is given by (\ref{eq:expmeanint}).
The second-order moment of the intensity is
\begin{equation}
%\EE \big[ I\big(z, \br +\frac{\bq}{2}\big) I\big(z, \br -\frac{\bq}{2}\big)\big]
\EE \big[ I\big(z, \bx_1\big) I\big(z, \bx_2\big)\big]
= \sum_{j,l=1}^2 \mu_{4,jl}(z ,\bx_1,\bx_2,\bx_1,\bx_2)  ,
\end{equation}
where the $\mu_{4,jl}$'s are defined by 
$$
\mu_{4,jl}(z,\bx_1,\bx_2,\by_1,\by_2 ) = \EE \big[ u_j(z,\bx_1) \overline{u_j(z,\by_1)}u_l(z,\bx_2) \overline{u_l(z,\by_2)}
 \big] .
$$
These moments satisfy the equation (\ref{eq:mu4}) and have the  initial conditions:
\begin{eqnarray}
\nonumber
&&
\mu_{4,jj}(z= 0 ,\bx_1,\bx_2,\by_1,\by_2) \\
\nonumber
&&= A_j^4
\exp\Big( -\frac{|\bx_1+\by_1|^2}{4 r_o^2} 
- \frac{|\bx_1-\by_1|^2}{4\rho_o^2}  -\frac{|\bx_2+\by_2|^2}{4 r_o^2} - \frac{|\bx_2-\by_2|^2}{4\rho_o^2}\Big) \\
&&\quad +A_j^4
\exp\Big( 
-\frac{|\bx_1+\by_2|^2}{4 r_o^2} 
- \frac{|\bx_1-\by_2|^2}{4\rho_o^2} 
-\frac{|\bx_2+\by_1|^2}{4 r_o^2} 
- \frac{|\bx_2-\by_1|^2}{4\rho_o^2}\Big) , \nonumber 
\end{eqnarray}
for $j=1,2$ and
\begin{eqnarray}
\nonumber
&&
\mu_{4,jl}(z= 0 ,\bx_1,\bx_2,\by_1,\by_2) \\
\nonumber
&&= A_j^2 A_l^2  
\exp\Big( -\frac{|\bx_1+\by_1|^2}{4 r_o^2} 
- \frac{|\bx_1-\by_1|^2}{4\rho_o^2}  -\frac{|\bx_2+\by_2|^2}{4 r_o^2} - \frac{|\bx_2-\by_2|^2}{4\rho_o^2}\Big) \\
&&\quad +A_j^2 A_l^2 \chi^2
\exp\Big( 
-\frac{|\bx_1+\by_2|^2}{4 r_o^2} 
- \frac{|\bx_1-\by_2|^2}{4\rho_1^2} 
-\frac{|\bx_2+\by_1|^2}{4 r_o^2} 
- \frac{|\bx_2-\by_1|^2}{4\rho_1^2}\Big) , \nonumber 
\end{eqnarray}
for $j\neq l$.
%Therefore we have
%$$
%\mu_{4,jj}(z= ,\bx_1,\bx_2,\by_1,\by_2)  = A_j^4 \mu_4 (z ,\bx_1,\bx_2,\by_1,\by_2) 
%$$
%and
%$$
%\mu_{4,jj}(z= ,\bx_1,\bx_2,\by_1,\by_2)  = A_j^4 \mu_4 (z ,\bx_1,\bx_2,\by_1,\by_2) 
%$$

Consequently, as shown in Appendix~\ref{sec:6},
in the scintillation regime
(which holds in the white-noise paraxial regime when additionally the correlation radius of the source is of the same order as the correlation radius of the medium, but the beam radius is much larger), the second-order moment of the intensity has the form
\begin{align}
\nonumber
\EE \big[ I\big(z, \br +\frac{\bq}{2}\big) I\big(z, \br -\frac{\bq}{2}\big)\big]
=&
\big(A_1^4+A_2^4\big) \big[ {\cal I}_{\rho_o} (2\br,{\bf 0})+{\cal J}_{\rho_o}(2\br,\bq)\big]\\
&
+
2A_1^2A_2^2 \big[ {\cal I}_{\rho_o} (2\br,{\bf 0})+\chi^2 {\cal J}_{\rho_1}(2\br,\bq)\big],
\end{align}
where ${\cal I}_{\rho_o}$ and ${\cal J}_{\rho_o}$ are defined by (\ref{eq:defIrho0}) and (\ref{eq:defJrho0}).
This can also be written as
\begin{align}
\nonumber
%\EE \big[ I\big(\frac{z}{\eps},\frac{\br}{\eps}+\frac{\bq}{2}\big) I\big(\frac{z}{\eps},\frac{\br}{\eps}-\frac{\bq}{2}\big)\big]
\EE \big[ I\big(z, \br +\frac{\bq}{2}\big) I\big(z, \br -\frac{\bq}{2}\big)\big]
=&
\big(A_1^4+A_2^4\big) \big[ {\cal H}_{\rho_o} (z,\br,{\bf 0})^2+|{\cal H}_{\rho_o}(z,\br,\bq)|^2\big]\\
&
+
2A_1^2A_2^2 \big[ {\cal H}_{\rho_o} (z,\br,{\bf 0})^2+\chi^2 |{\cal H}_{\rho_1}(z,\br,\bq)|^2 \big],
\end{align}
or equivalently (\ref{eq:expcovint}).
This shows that the field satisfies the Gaussian rule for the fourth-order moment in the scintillation regime:
\begin{align}
\nonumber
&
%\EE \big[ I\big(\frac{z}{\eps},\frac{\br}{\eps}+\frac{\bq}{2}\big) I\big(\frac{z}{\eps},\frac{\br}{\eps}-\frac{\bq}{2}\big)\big]\\
\EE \big[ I\big(z, \br +\frac{\bq}{2}\big) I\big(z, \br -\frac{\bq}{2}\big)\big]\\
\nonumber &=
\sum_{j,l=1}^2 
\EE \big[ u_j\big(z,\br +\frac{\bq}{2}\big) 
\overline{u_j}\big(z, \br +\frac{\bq}{2}\big)\big]
\EE \big[ u_l\big(z, \br -\frac{\bq}{2}\big) 
\overline{u_l}\big(z, \br -\frac{\bq}{2}\big)\big]
%\EE \big[ u_j\big(\frac{z}{\eps},\frac{\br}{\eps}+\frac{\bq}{2}\big) 
%\overline{u_j}\big(\frac{z}{\eps},\frac{\br}{\eps}+\frac{\bq}{2}\big)\big]
%\EE \big[ u_l\big(\frac{z}{\eps},\frac{\br}{\eps}-\frac{\bq}{2}\big) 
%\overline{u_l}\big(\frac{z}{\eps},\frac{\br}{\eps}-\frac{\bq}{2}\big)\big]
\\
&\quad +
\sum_{j,l=1}^2 
\EE \big[ u_j\big(z, \br+\frac{\bq}{2}\big) 
\overline{u_l}\big(z, \br -\frac{\bq}{2}\big)\big]
\EE \big[ u_l\big(z, \br -\frac{\bq}{2}\big) 
\overline{u_j}\big(z, \br +\frac{\bq}{2}\big)\big]
%\EE \big[ u_j\big(\frac{z}{\eps},\frac{\br}{\eps}+\frac{\bq}{2}\big) 
%\overline{u_l}\big(\frac{z}{\eps},\frac{\br}{\eps}-\frac{\bq}{2}\big)\big]
%\EE \big[ u_l\big(\frac{z}{\eps},\frac{\br}{\eps}-\frac{\bq}{2}\big) 
%\overline{u_j}\big(\frac{z}{\eps},\frac{\br}{\eps}+\frac{\bq}{2}\big)\big]
.
\end{align}
 
The quasi-Gaussianity property and intensity covariance expression that we have just identified 
are useful in various applications. 
For instance in applications to imaging based on wave field coherence
the signal-to-noise ratio will in general depend on fourth-order wave field moments \cite{borcea11,book16}. 
These moments can then be related to the mutual coherence function for the wave field
via the theory presented in this paper.  
 We discuss next an application to the 
inverse source problem for partially coherent beam sources  in complex media
using information that can be extracted from the mean intensity covariance.  
     
\section{Estimation of  Partially Coherent Electromagnetic Sources}
\label{sec:app}

 We consider  the problem of characterizing a partially coherent source.  
 The goal is to determine the parameters of a Gauss-Schell source from the measurements of the intensity after a propagation distance $z$. 
 Note that here we base  the estimate on measurements of  intensity only while 
 in for instance in \cite{ip} the estimate is based on coherence of the measurements 
 of the field itself.   In \cite{liu} the authors use a cross phase in the source 
 for stable transmission of the coherence pattern in the source field. This then allows
 for transmission of information. In order to relate the field coherence to coherence in the  intensity the authors use a Gaussian approximation for the field. It follows from the analysis 
 in our paper that such an approximations is valid in the scintillation regime.  
 We remark that in the case with active multifrequency imaging one can obtain partial phase information from an appropriate illumination
 stategy and use of a polarization identity \cite{int-im}. 
 Here we consider passive single-frequency intensity based 
 imaging and no phase information is available.  
   
 In the source estimation context considered here we assume  the scintillation regime, moreover,  we use the strongly scattering 
 approximation studied in Appendix \ref{sec:5} (which holds when $k_o^2 C({\bf 0}) z  \gg  1$ and $C$ is smooth).
 We then find
\begin{equation}
\EE \big[ I(z,\br) \big] =(A_1^2+A_2^2) \frac{r_o^2}{ R^2(z;\rho_o) }
\exp \Big( - \frac{|\br|^2}{R^2(z;\rho_o)} \Big) ,
\label{eq:meanint}
\end{equation}
where the beam radius is
\begin{equation} 
R^2(z;\rho_o) = 
r_o^2 \Big( 1+\frac{\bar{\gamma}_2 z^3}{3 r_o^2}+\frac{z^2}{k_o^2 r_o^2 \rho_o^2}\Big)
%= r_o^2 + \frac{\bar{\gamma}_2 z^3}{3}+\frac{ z^2}{k_o^2 \rho_o^2}
  ,
\label{eq:defRz}
\end{equation}
with $\bar{\gamma}_2= - \Delta_\bx C({\bf 0}) /4$ being a parameter that governs the
strength of random lateral scattering.  In this expression the original beam radius
is $r_o$, the third term in the right-hand side  gives the spreading due to diffraction ${\cal O}(z)$  in the  homogeneous medium case.   The 
second term gives  the anomalous spreading ${\cal O}(z^{3/2})$ due to random scattering.  This  expression for the beam spreading
was also derived in \cite{gburw} using the Huygens-Fresnel principle.   
 
The covariance intensity function is
 \ba
 \nonumber
&& {\rm Cov}\big(  I\big(z, \br +\frac{\bq}{2}\big) , I\big(z, \br -\frac{\bq}{2}\big)\big) = \\&& \nonumber
\big(A_1^4+A_2^4\big) 
\frac{r_o^4}{R(z;\rho_o)^4} 
  \exp\Big( - \frac{2|\br|^2}{R^2(z;\rho_o)} 
 -\frac{|\bq|^2}{2 \rho^2(z;\rho_o)}  \Big) \\
&& +
2A_1^2A_2^2 \chi^2 
\frac{r_o^4}{R(z;\rho_1)^4} 
  \exp\Big( - \frac{2|\br|^2}{R^2(z;\rho_1)} 
 -\frac{|\bq|^2}{2 \rho^2(z;\rho_1)}  \Big), \label{eq:meanco}
 \ea
   where the correlation radius of the beam is
\begin{equation}
\rho^2(z;\rho_o) = \rho_o^2 \frac{1+ \frac{\bar{\gamma}_2 z^3}{3 r_o^2}+\frac{ z^2}{k_o^2 \rho_o^2 r_o^2}}{
1+k_o^2 \rho_o^2 \bar{\gamma}_2 z \big( 1+ \frac{\bar{\gamma}_2 z^3}{12 r_o^2}+\frac{ z^2}{3 k_o^2 \rho_o^2 r_o^2}\big)} .  
\label{eq:defrhoz}
\end{equation}
In the homogeneous medium case we have  $\rho(z;\rho_o) / R(z;\rho_o) = \rho_o  /r_o$
and the lateral correlation radius increases with the propagation distance  while
in the random case $\rho(z;\rho_o) = {\cal O}(z^{-1/2})$  and 
the lateral correlation radius decreases with the propagation distance 
due to random scattering.   

The structure of the intensity covariance in lateral and range coordinates 
can be used for imaging of the partially coherent source. 
 Here we consider measurements  at one range $z$ only. 
Note that the expressions in    (\ref{eq:meanint}) and (\ref{eq:meanco}) are 
computed based  on averaging with respect to the statistics of both the source and of the medium.   
We assume that these means can be identified with a high signal-to-noise ratio. 
This is the case if both the partially coherent source and the medium fluctuate in time
and we average the measurements over a time interval that is long compared to the turnover
times of the medium  and of the source. 
%A detailed analysis of the case when the detector 
%averaging time is not large compared to the  turnover times is in \cite{gs-josa21}.  
In cases when the averaging is not efficient or the 
medium is stationary (time-independent) it may be necessary 
with some form of filtering to enhance statistical stability \cite{wrcm-bpt}.  

In the statistically stable case with long time averaging at the detector 
the observation of the mean intensity and the intensity covariance function makes it possible to extract the beam radii 
$R(z;\rho_o)$, $R(z;\rho_1)$, correlation radii $\rho(z;\rho_o)$, $\rho(z;\rho_1)$ and the 
intensity amplitude $(A_1^2+A_2^2)r_o^2$. 
If $\rho_o \neq \rho_1$, then we can also extract $(A_1^4+A_2^4)r_o^4$ and $2A_1^2 A_2^2\chi^2 r_o^4$.
If $\rho_o=\rho_1$, then we can only extract the sum $(A_1^4+A_2^4+2A_1^2 A_2^2\chi^2) r_o^4$. 

Given the values of $z$ and $\bar{\gamma}_2$,
we can then estimate the beam radius $r_o$ of the source, the correlation radii $\rho_o$ and $\rho_1$, the total intensity $A_1^2+A_2^2$.
%and the square correlation degree $\chi^2$ since
%$\chi^2= [(A_1^4+A_2^4+2A_1^2 A_2^2\chi^2) r_o^4 ] / [ (A_1^2+A_2^2)^2r_o^4 -(A_1^4+A_2^4)r_o^4 %]$.
If $\rho_o \neq \rho_1$ 
 we can also extract the polarization degree $\chi$ and the ellipticity 
 $e^2= (A_2^2-A_1^2)^2 /(A_1^2+A_2^2)^2$. 
Indeed, if we introduce the three following quantities that can be extracted from data 
$Y_1=(a_1^2+a_2^2)$ and $Y_2=a_1^4+a_2^4$, $Y_3=2a_1^2 a_2^2 \chi^2$
with $a_j=A_j r_o$,
  we have
  \begin{equation}
    e^2=  \frac{ (a_1^2-a_2^2)^2 }{ (a_1^2+a_2^2)^2 } = 2\frac{Y_2}{Y_1^2}-1, 
    \quad \quad \chi^2=\frac{ Y_3}{Y_1^2-Y_2} .
   \end{equation}
 Then, with an estimate of  $r_o$ we can identify $A_1$ and $A_2$.  
 
The estimation is possible and reliable provided the propagation distance is not too large, i.e., $z$ should not be much larger than 
\ban
\max\big( (r_o^2/\bar{\gamma}_2)^{1/3}, 1/(\bar{\gamma}_2 k_o^2 \rho_o^2), 1/\bar{\gamma}_2 \big),
\ean
 because the statistics of the beam 
(beam radius and correlation radius) then becomes essentially independent of the initial values $r_o$ and $\rho_o$ as shown in  Appendix \ref{sec:5}.
We remark also that when $\bar{\gamma}_2 z \gg 1$ the paraxial approximation is not valid anymore, as 
shown in Appendix \ref{sec:5}, so it is not possible to use this approach.
Note that that  in this regime  one can then use 
a least-squares misfit approach \cite{ip} to estimate 
$r_o, \rho_o$ and $\rho_1$ from $R(z;\rho_o)$,  $R(z;\rho_1)$,  $\rho(z;\rho_o)$
and $\rho(z;\rho_1)$.

% We comment next on estimation of $r_o$ and $\rho_o$ and consider a regime when the
%propagation distance is such that
%$\bar{\gamma}_2 z  \ll 1$.
%We then have
%\begin{eqnarray*}
%&& r_o^2 =  R^2(z;\rho_o) - \frac{z^2}{\rho^2(z;\rho_o) k_o^2}   
%+ \frac{2 \bar{\gamma}_2 z^3}{3} -\frac{z^4}{ R^2(z;\rho_o) \rho^4(z;\rho_o) k_o^4 }
%+ {\mathcal O}((\bar{\gamma}_2 z)^5),\\
%&& \rho_o^2 =   \rho^2(z;\rho_o) \Big(
%    1+ \rho^2(z;\rho_o) k_o^2 (\bar{\gamma}_2 z)
%    + \big( \rho^4(z;\rho_o) k_o^4 - (\bar{\gamma}_2 k_o \rho(z;\rho_o)  R(z;\rho_o))^{-2} \big) (\bar{\gamma}_2 z)^2  \\ && \hbox{} 
%    - \frac{(\bar{\gamma}_2 z)^3 }{\bar{\gamma}_2^2 R^2(z;\rho_o)}
%    + \big( (k_o \bar{\gamma}_2 \rho(z;\rho_o)  R(z;\rho_o) )^{-4} - \frac{ 15 k_o^2  \rho^2(z;\rho_o)}{12 \bar{\gamma}^2_2 R^2(z;\rho_o) } \big) (\bar{\gamma}_2 z)^4
%\Big)+{\mathcal O}((\bar{\gamma}_2 z)^5) .    
%\end{eqnarray*}
%In the general case with $\bar{\gamma}_2 z={\mathcal O}(1)$ one can use 
%a least-squares misfit approach \cite{ip} to estimate 
%$r_o, \rho_o$ and $\rho_1$ from $R(z;\rho_o)$,  $R(z;\rho_1)$,  $\rho(z;\rho_o)$
%and $\rho(z;\rho_1)$.
%When $\bar{\gamma}_2 z \gg 1$ the paraxial approximation is not valid anymore, as 
%shown in Appendix \ref{sec:5}, so it is not possible to use this approach anymore.

 \section{Conclusions} 
 
 We have considered time-harmonic electromagnetic wave propagation from partially coherent sources
 in random media. In many applications of waves it is of interest to evaluate
 the fourth moment of the wave field. Here we present a theory that allows us to describe
 such moments and we focus on the specific fourth moments corresponding to 
 the intensity covariance. 
 We present here such a description for polarized waves.  
 The results follow from the It\^o-Schr\"odinger equation
 for  the wave field valid in the white-noise paraxial regime. 
 An important aspect of these  It\^o-Schr\"odinger equations
   is that  the equations describing the evolutions of the transversely polarized modes
 are driven by the same Brownian motion, however, 
 such that they are dynamically uncoupled.   
 The explicit expressions for the fourth moments are derived in a subsequent scaling
 regime that we denote the scintillation regime.  
 An important aspect of the 
 fourth moment analysis is the proof of the quasi-Gaussian property
  which means that the fourth moments can be obtained from the second
  wave field moments  as if the field had Gaussian statistics. 
  We note  that this property 
  holds true even if the wave field is partially coherent. We also describe an application
  to the inverse source problem using information extracted from the observed intensity covariance.
 We moreover give explicit expressions for the decorrelation and spreading scales deriving from the mutual coherence function   for probing through strong clutter, 
  these scales characterize the statistical structure of the wave field in view of the quasi-Gaussian property (up to fourth order).

\appendix

\section{The Scintillation  Regime for the Electromagnetic  Waves}
\label{app:a}%
 We discuss here the white-noise paraxial scaling regime that leads to the 
 It\^o-Schr\"odinger equation  in (\ref{eq:model2}). We refer to \cite{59}
 for the full derivation. 
The electromagnetic wave equations have the form (\ref{eq:0a}-\ref{eq:0d})
  with  the dielectric permittivity  $\epsilon$ and
the magnetic permeability $\mu$ of the medium modeled by (\ref{eq:med10}-\ref{eq:med20}).
 We  denote $\vec{\itbf E}   = ( {E}_j)_{j=1,2,3}$ 
and $ \vec{\itbf H} =({H}_j)_{j=1,2,3}$.
The four-dimensional vector $( {E}_1, {H}_2, {E}_{2},{H}_{1})$ then satisfies a closed system as shown in \cite{59}.
 Let  $c_o = \mu_o^{-1/2} \epsilon_o^{-1/2}$ 
and $\zeta_o= \mu_o^{1/2} \epsilon_o^{-1/2}$ be  the homogeneous propagation speed
and impedance, then we introduce the decomposition 
\begin{eqnarray*}
{E}_1 
(z,\bx)
 &=& \zeta_o^{\frac{1}{2}}
\big( 
{a}_1 (z,\bx) e^{i \frac{\omega z}{c_o }}
+
{b}_1 (z,\bx) e^{-i \frac{\omega z}{c_o }}
\big) ,\\
{H}_2 
(z,\bx) 
&=&  \zeta_o^{-\frac{1}{2}}
\big( 
{a}_1 (z,\bx) e^{i \frac{\omega z}{c_o }}
-
{b}_1 (z,\bx) e^{-i \frac{\omega z}{c_o }}
\big) ,\\
{E}_2
(z,\bx) 
&=& \zeta_o^{\frac{1}{2}}
\big( 
{a}_2 (z,\bx) e^{i \frac{\omega z}{c_o }}
+
{b}_2 (z,\bx) e^{-i \frac{\omega z}{c_o }}
\big) ,\\
{H}_1 
(z,\bx) 
&=&  \zeta_o^{-\frac{1}{2}}
\big( -
{a}_2 (z,\bx) e^{i \frac{\omega z}{c_o }}
+
{b}_2 (z,\bx) e^{-i \frac{\omega z}{c_o }}
\big) .
\end{eqnarray*}
Here, $a_j,b_j, j=1,2$ are coefficients of locally forward and backward (in $z$)
propagating plane waves. In the case of a homogeneous medium with $m_\epst\equiv 0, m_\mu\equiv 0$
this gives an exact decomposition into forward and backward plane waves with constant coefficients,  while  with
random  medium fluctuations the coefficients $a_j$ and $b_j$ satisfy coupled equations.

Let $\sigma$ be the standard deviation of the fluctuations of the medium.   Moreover, assume  that the random fluctuations in the index of
refraction are isotropic and denote by $\ell_{\rm c}$ the correlation length of the fluctuations, $\lambda_o$ 
the carrier wavelength (equal to $2\pi /k_o$, $k_o=\omega_o/c_o$), $L$ the typical propagation distance, $\rho_o$ the correlation radius of the source, and $r_o$
the radius of the initial transverse beam-source.
In this framework the variance $C({\bf 0})$ of the Brownian field in the It\^o-Schr\"odinger equation
(\ref{eq:model2}) is of order $\sigma^2 \ell_{\rm c}$ and the transverse scale of variation of the covariance function
$C(\bx)$ in (\ref{def:C2}) is of order $\ell_{\rm c}$.

We next  discuss the scintillation regime in more detail.   
First,  we consider the primary scaling (white-noise paraxial regime) that leads to
the It\^o-Schr\"odinger equation (\ref{eq:model2}), when the propagation distance is much larger than  the correlation length of the medium, the correlation radius of the source and the beam radius, which are themselves much larger
than the wavelength,
moreover, the medium fluctuations are small. 
Explicitly, we assume the primary scaling  when
\begin{eqnarray*}
   \frac{\rho_o}{\ell_{\rm c}} \sim 1  \, ,   \quad\quad
  \frac{r_o}{\ell_{\rm c}} \sim 1  \, ,   \quad\quad
      \frac{L}{\ell_{\rm c}} \sim   \alpha^{-1}\, ,    \quad \quad
     \frac{\lambda_o}{\ell_{\rm c}}    \sim    \alpha\,   ,     \quad  \quad
     \sigma^2 \sim \alpha^3 \,  , 
\end{eqnarray*}
where $\alpha$ is a small dimensionless parameter.
We introduce dimensionless coordinates by:
\begin{eqnarray*}
&&\bx = { \ell_{\rm c} }  { \bx' }  , \quad  \quad  
%z =  { \ell_{\rm c} } z', \quad\quad  
z =  L z', \quad\quad 
k_o = \frac{k_o' }{  \ell_{\rm c} \alpha }, \\
&& m_\eps ( L z',  \ell_{\rm c} \bx') = \alpha^{3/2}  m_\eps' \left(z' ,  \bx'  \right)  , \quad \quad  
m_\mu ( L z',  \ell_{\rm c} \bx') = \alpha^{3/2}  m_\mu' \left(z' ,  \bx'  \right)   .
\end{eqnarray*}
We look for the behavior of the coefficient 
%$u_j(z',\bx') = a_j\big( \frac{z' \ell_{\rm c}}{\alpha},\bx' \ell_{\rm c}\big)$
$u_j(z',\bx') = a_j\big( z' L ,\bx' \ell_{\rm c}\big)$
for long propagation distances of the order of $\alpha^{-1}$.
We obtain a Schr\"odinger-type equation 
in which the potential fluctuates in $z'$ on the  
scale $\alpha$  and is of amplitude
$ \alpha^{-1/2}$. 
This diffusion approximation scaling gives the Brownian field and the model (\ref{eq:model2}).
As follows from our analysis in \cite{59},
the backward propagating wave components 
$b_j$, $j=1,2$ are small compared to the forward propagating wave components 
$a_j$, $j=1,2$ in this forward beam propagation regime,
We remark also that the  local propagation speed is
$$
c  = \frac{1}{\sqrt{ \mu' \epst'}} = c_o \big[ 1 - \alpha^{3/2} \frac{m'_\mu+m'_\epst}{2} +O(\alpha^3)\big],
$$ 
and the local impedance
$$
    \zeta = \sqrt{ \mu' / \epst'} = \zeta_o \big[ 1 +\alpha^{3/2} \frac{m'_\mu-m'_\epst}{2}  +O(\alpha^3) \big].
$$ 
In view of (\ref{def:C2}) it then follows that the effective Brownian field
is determined by the fluctuations of the local propagation speed, but not by the fluctuations of the local impedance.

In Appendix \ref{sec:6} we address the subsequent scaling regime in which the correlation 
length of the medium $\ell_{\rm c}$ and the correlation radius $\rho_o$ of the source are much smaller than the beam radius $r_o$ of the source,
moreover, the medium fluctuations are weak and the beam propagates deep
into the medium. We  then get the modified scaling picture
\begin{equation}
\label{eq:scaling app}
    \frac{\rho_o}{\ell_{\rm c}} \sim  1  \, ,  \quad\quad
        \frac{r_o}{\ell_{\rm c}} \sim  \eps^{-1}  \, ,  \quad\quad
     \frac{L}{\ell_{\rm c}} \sim   \alpha^{-1} \eps^{-1} \, ,   \quad\quad
     \frac{\lambda_o}{\ell_{\rm c}} \sim \alpha   \,   ,    \quad\quad
     \sigma^2 \sim \alpha^3 \eps \,      ,
\end{equation}
and we assume $  \alpha \ll \eps \ll 1$.
This means that the paraxial white-noise limit $\alpha \to 0$ is taken first, 
and we find
$$ 
2ik_o d {u}_j^\eps      
    +\Delta_{\bx}   {u}_j^\eps \, dz
   +  k_o^2   {u}_j^\eps   \circ  d{B}^\eps(z,\bx) 
 =0 ,
$$
where the Brownian field ${B}^\eps$ has covariance $C^\eps$.
%where (relatively to the correlation length of the medium), the initial correlation radius of the source is of order one, the initial beam radius $r_o$ is of  order $\eps^{-1}$,  
%the variance $C^\eps({\bf 0})$ of the Brownian field $B^\eps$ is of order $\eps$,
%and the propagation distance $L$ is of order $\eps^{-1}$.  
Then the limit  $\eps\to 0$ is applied, corresponding to the scintillation regime. 
In the scintillation regime (\ref{eq:scaling app}) the effective strength $k_o^2 C^\eps({\bf 0}) L$ of the 
Brownian field is of order one since $  \sigma^2 \ell_{\rm c} L/\lambda_o^2   \sim 1$.
We also have that  $L \lambda_o/ r_o^2$ is of order $\eps$.
That is,   the typical propagation distance is smaller than the Rayleigh length associated to a coherent beam with radius $r_o$.
Here the Rayleigh length corresponds to the distance when
the transverse radius of a coherent beam with radius $r_o$ has roughly  doubled by diffraction in the homogeneous medium case and it  is given by $r_o^2/\lambda_o$.
The typical propagation distance is, however, of the same order as the Rayleigh length associated to a partially coherent beam with beam radius $r_o$ and correlation radius $\rho_o$, which is given by $r_o \rho_o/\lambda_o$ \cite{gar00}.
The scintillation regime is, therefore, a regime where diffractive and random effects are both effective and their combination results in non-trivial effects.
In this regime we are also able to derive explicit expressions for the fourth 
moment, see Appendix \ref{sec:6}.

%     
%In Appendix  \ref{sec:6}  we consider the white noise paraxial regime in more
%detail  and in particular there consider a subsequent scaling  that we refer 
%to as  the white-noise scintillation regime. 
% This corresponds to a situation in which the relative intensity fluctuations are
%of order one and it is an important regime to capture from the physical viewpoint.
%

\section{The Mean Polarized-Wigner Transform  for a Partially Coherent Beam} 
 \label{sec:4}%
 We discuss here the Wigner transform which is convenient
 in order to describe second order field moments. 
 Let $j,l\in \{1,2\}$.
The mean Wigner transform is defined by 
\begin{equation}
{W}_{\rm m} ( z,\br,\bxi ) := 
\int_{\RR^2}  
\exp \big( - i  \bxi \cdot \bq  \big)
\EE \left[ {u_j}\big(z,  \br+\frac{\bq}{2} \big)
 \overline{u}_l \big( z,   \br-\frac{\bq}{2}\big) \right]  d \bq  .
\end{equation}
In view of Eq.~(\ref{eq:mu2}) it satisfies the closed system
\begin{equation}
\label{systemWT2rapid}
\frac{\partial {W}_{\rm m} }{\partial z} + \frac{1}{k_o}
{ \bxi}\cdot \nabla_{\br} W_{\rm m}
=\frac{k_o^2}{4 (2\pi)^2}
\int_{\RR^2}  \hat{C}( \bk) \Big[ 
W_{\rm m} (   \bxi - \bk   ) 
 - W_{\rm m}(  \bxi ) 
\Big]  d \bk ,
\end{equation}
starting from ${W}_{\rm m}( z= 0 ,\br,\bxi) =W_{{\rm m}0}(\br,\bxi)$, which is the mean Wigner transform of the source~$(f_j,f_l)$:
$$
W_{{\rm m}0}(\br,\bxi) := \int_{\RR^2}  \exp \big(- i\bxi \cdot \bq \big)
\EE \Big[ f_j \big( \br + \frac{\bq}{2}\big) \overline{f}_l \big( \br - \frac{\bq}{2}\big) \Big] d\bq.
$$
 The transport equation (\ref{systemWT2rapid}) can be solved and we find
\begin{align}
\nonumber
W_{\rm m}(z,\br,\bxi)=&  \frac{1}{(2\pi)^{2}} \iint_{\RR^2\times \RR^2}
\exp \Big( i  \bzeta \cdot \big( \br -\bxi \frac{z}{k_o} \big) - i  \bxi  \cdot \bq \Big)\hat{W}_{{\rm m}0}\big(\bzeta,\bq  \big) 
\\
& \times  \exp \Big( \frac{k_o^2}{4} \int_0^z {C} \big( \bq + \bzeta \frac{z'}{k_o} \big)-{C}(
 {\bf 0}) d z' \Big)
 d\bzeta d \bq ,
\label{solWT1b}
\end{align}
where $\hat{W}_{{\rm m}0}$ is defined in terms of  the source~$(f_j,f_l)$ as:
\begin{equation}
\label{eq:part}
\hat{W}_{{\rm m}0}(\bzeta,\bq)  = \int_{\RR^2}  
\exp \big( - i  \bzeta \cdot \br  \big)
\EE \Big[  f_j\big(   \br+\frac{\bq}{2} \big)
 \overline{f}_l \big(     \br-\frac{\bq}{2}\big)  \Big]  d \br  .
\end{equation}
The mean Wigner transform gives an equation for the 
mutual coherence function and we next discuss this in
a situation with strong scattering.

\section{The Mutual Coherence Function in the Strongly Scattering Regime}
\label{sec:5}%
We consider a Gauss-Schell model for the source, which is a field with Gaussian statistics, mean zero, and covariance function 
(\ref{eq:ic}).
By taking the inverse Fourier transform of the mean Wigner transform, we find that the covariance function of the transmitted field has 
the form (\ref{eq:covfield}).
We discuss next this mutual coherence function in more detail
and to find explicit expressions we assume in the rest  of this section
that scattering is strong and smooth, in the sense that 
 \begin{align}
\label{def:gam}
& k_o^2 C({\bf 0}) z  \gg  1  , \\ 
& C(\bx) = C({\bf 0})  - \bar{\gamma}_2|\bx|^2 +o(|\bx|^2) .
\label{def:gam2}
\end{align}
From (\ref{eq:model2}) written in It\^o form \cite{oks} 
it follows that  the scattering mean free path
$\ell_{\rm mfp}$ (that is the typical propagation distance 
over which a coherent wave becomes incoherent) is
\begin{equation}
\ell_{\rm mfp}^{-1}  = \frac{  k_o^2  C({\bf 0}) }{8}=\frac{  k_o^2  \sigma^2 \ell_z }{8} .
\end{equation}
Thus,  in the regime (\ref{def:gam}) the propagation distance is large compared to the scattering mean free
path.  
Note moreover that $\bar{\gamma}_2$ can be interpreted as 
\begin{equation}
\bar{\gamma}_2 = \frac{\sigma^2 \ell_z}{\ell_\perp^2}
\end{equation}
where $ \sigma^2$ is the variance of medium fluctuations as above,
 $\ell_z$ is the longitudinal correlation length of the medium (such that $C({\bf 0})=\sigma^2 \ell_z$), and $\ell_\perp$ is its transverse correlation radius of the medium defined by:
\begin{equation}
\ell_\perp^{-2} = - \frac{\Delta C({\bf 0})}{4C({\bf 0})} .
\end{equation}
If the medium is isotropic (as assumed in Appendix \ref{app:a}),
for instance such that 
$\EE[(m_\epst+m_\mu)(\bx'+  z,\bx'+z)(m_\epst+m_\mu)(\bx',z')] = \sigma^2 \exp( - |\bx|^2/\ell_{\rm c}^2-z^2/\ell_{\rm c}^2)$,
  then we have $\ell_z=\sqrt{\pi}\ell_{\rm c}$ and $\ell_\perp=\ell_{\rm c}$.
%as in the previous appendix.  

If (\ref{def:gam}-\ref{def:gam2}) hold, then  the mean intensity is given by (\ref{eq:meanint}).
This is found via a Gaussian calculation after inserting (\ref{def:gam}) in (\ref{def:Hq}). 
In this expression we can identify the original beam radius $r_o^2$, the spreading due to diffraction $\frac{ z^2}{k_o^2 \rho_o^2} $, and the spreading due
to random scattering $ \frac{\bar{\gamma}_2 z^3}{3}$.
The covariance function of the field (or mutual coherence function) is given by
\ban
&&\EE \left[ {u}_j\big(z,  \br+\frac{\bq}{2} \big)
 \overline{u_j } \big( z,   \br-\frac{\bq}{2}\big) \right] 
 = \\ && 
A_j^{2}  \frac{r_o^2}{R(z;\rho_o)^2}  \exp\Big( - \frac{|\br|^2}{R^2(z;\rho_o)} 
 -\frac{|\bq|^2}{4 \rho^2(z;\rho_o)}  + i  \theta(z;\rho_o) \br\cdot\bq \Big),
\ean
  where the correlation radius  of the beam, $\rho(z;\rho_o)$, 
  is given by (\ref{eq:defrhoz}), the beam radius, 
  $R(z;\rho_o)$, by (\ref{eq:defRz}) 
and
\begin{equation}
\label{eq:defthetaz}
\theta (z;\rho_o) = \frac{\frac{z}{k_o \rho_o^2} + \frac{k_o\bar{\gamma}_2 z^2}{2} }{r_o^2  +\frac{\bar{\gamma}_2 z^3}{3}+\frac{ z^2}{k_o^2 \rho_o^2}  }.
\end{equation}
If the source is coherent $\rho_o=r_o$, then we recover the classical result obtained in \cite{gar09a}, while 
if the medium is homogeneous and the source is partially coherent $\rho_o <  r_o$,
then we recover the result obtained in \cite{gar00}. 
Note that if the medium  is homogeneous then 
$ 1/\sqrt{\theta(z;\rho_o)}  \ll  \rho(z;\rho_o) \leq R(z;\rho_o), ~ \hbox{as}~ z\to \infty$,
 while 
in the random case   $\rho(z;\rho_o) \ll 1/\sqrt{\theta(z;\rho_o)}  \ll  R(z;\rho_o), ~ \hbox{as}~ z\to \infty$.
In fact in both cases we have $1/\sqrt{\theta} \sim  \sqrt{\lambda_o z}$, up to a constant.    Thus, the coherent phase modulation 
is slow relative to the field decorrelation scale for deep probing
in the random medium.

%If one can observe the covariance function (more realistically, if we can observe the real part of the covariance function by looking at the mean intensities 
%$\EE\big[ \big|{u}\big(z,  \br+\frac{\bq}{2} \big) + {u} \big( z,   \br-\frac{\bq}{2}\big)\big|^2\big]- \EE\big[ \big|{u}\big(z,  \br+\frac{\bq}{2} \big) \big|^2\big]-\EE\big[\big|   {u} \big( z,   \br-\frac{\bq}{2}\big)\big|^2\big]$), then it is possible to estimate $R(z)$, $\rho(z)$ and $\theta(z)$. 
%Given $z$,
%we can then invert the relations (\ref{eq:defRz}-\ref{eq:defrhoz}-\ref{eq:defthetaz}) to get estimates of $\bar{\gamma}_2$, $r_o$ and $\rho_o$, which gives the characteristics of the partially coherent source.

For large propagation distance so that the spreading due to the random medium dominates,
$z \gg \max\big( (r_o^2/\bar{\gamma}_2)^{1/3}, 1/(\bar{\gamma}_2 k_o^2 \rho_o^2) \big)$, we have:
\begin{align}
R(z;\rho_o)  &\sim \cA(z) :=\sqrt{\frac{ \bar{\gamma}_2 z^3 }{3} }       , \\
\rho(z;\rho_o)  &\sim\rho_{\rm tr}(z) := \frac{ 1}{k_o \sqrt{\bar{\gamma}_2 z }}   , \\
\theta(z;\rho_o) &\sim  \frac{3 k_o}{ 2 z} .
 \end{align}
 Note that these parameters are independent of the parameters $r_o$ and $\rho_o$  of  the 
 partially coherent source,  so that information about the source is ``forgotten''  in the case of deep probing.
We refer to the parameters  $ \cA,  \rho_{\rm tr}$ as the time reversal aperture and
resolution respectively. We note that we have the Rayleigh  resolution relation
\ba 
    \rho_{\rm tr}(L)  = \frac{ \lambda_o L}{ \cA(L) } .
\ea
where   $\rho_{\rm tr}$  corresponds to the refocusing
 resolution one obtains when a point source emits a wave which is captured 
 on a time reversal mirror at distance $L$ and reemitted (after time reversal) toward the source. Then it will refocus at the original source location
  with a resolution of the order of the lateral correlation range $ \rho_{\rm tr}(L)$ 
  essentially independently of the actual physical aperture \cite{gar17}.
  %NB , see Figure \ref{fig:1}. 
  This can be understood in that 
  the propagator of the transmitted  wave  decorrelates (laterally) on this scale and
  the refocused wave is essentially the convolution of the propagator with itself.

We remark that the field correlation radius $\rho(z;\rho_o)$ is important in determining statistical stability.
If we average a field quantity over an aperture then the signal-to-noise ratio will in general depend on the ratio 
of  the aperture to the field correlation radius.
 We remark finally that we have $\rho_{\rm tr} \approx  \lambda_o$ when $ z\approx 1/{ \bar{\gamma}_2} $, 
which means that the paraxial approximation is not valid beyond this propagation distance.

\section{The Intensity Covariance Function}
\label{sec:6}%
 
In this appendix we derive the expression for the intensity covariance function 
in the scintillation regime.  
We start by introducing  
\begin{equation}
\label{def:generalmoment}
\mu_{4,j_1l_1j_2l_2}(z,\bx_1,\bx_2,\by_1,\by_2 ) = \EE \big[ u_{j_1}(z,\bx_1) \overline{u_{l_1}(z,\by_1)}u_{j_2}(z,\bx_2) \overline{u_{l_2}(z,\by_2)}
 \big] .
\end{equation}
We are interested in the second-order moment of the intensity:
\begin{equation}
 \EE \big[ I(z,\bx_1) I(z,\bx_2)  \big] = 
 \sum_{j_1,j_2=1}^2
 \mu_{4,j_1j_1j_2j_2} (z,\bx_1,\bx_2,\bx_1,\bx_2 ) .
\end{equation}
We find using Eq.~(\ref{eq:model2}) that the general fourth-order moment $\mu_{4,j_1l_1j_2l_2}$  satisfies the equation
\ba\nonumber && 
\frac{\partial \mu_{4,j_1l_1j_2l_2}}{\partial z} = \frac{i}{2k_o}  \Big( \Delta_{\bx_1}+ \Delta_{\bx_2} 
-  \Delta_{\by_1} -  \Delta_{\by_2} \Big) \mu_{4,j_1l_1j_2l_2} \\ && \hbox{} + \frac{k_o^2}{4} {U}_{4} \big(  \bx_1,\bx_2,  \by_1,\by_2\big)
\mu_{4,j_1l_1j_2l_2}  ,
 \label{eq:mu4}
 \ea
with the generalized potential
\begin{eqnarray}
\nonumber
{U}_{4}\big( \bx_1,\bx_2,  \by_1,\by_2\big)  
&=& 
{C}(\bx_1-\by_1) 
+
{C}(\bx_2-\by_2) 
+
  {C}(\bx_1-\by_2) 
+
  {C}(\bx_2-\by_1) 
\\
&&-
  {C}(\bx_1-\bx_2) 
-
  {C}(\by_1-\by_2) 
-2 {C}({\bf 0})  ,
\end{eqnarray}
and the initial condition:
$$
\mu_{4,j_1l_1j_2l_2}(z= 0 ,\bx_1,\bx_2,\by_1,\by_2) = 
\EE \big[f_{j_1}(\bx_1) \overline{f_{l_1}(\by_1)} f_{j_2}(\bx_2) \overline{f_{l_2}(\by_2)} \big] .
$$
Using the Gaussian property of the source, the initial condition for the fourth-order moment is:
\begin{eqnarray}
\nonumber
&&
\mu_{4,j_1l_1j_2l_2}(z= 0 ,\bx_1,\bx_2,\by_1,\by_2) \\
\nonumber
&&= 
\EE\big[ f_{j_1} (\bx_1)\overline{f_{l_1}}(\by_1) \big]
\EE\big[ f_{j_2}(\bx_2)\overline{f_{l_2}}(\by_2) \big]
+
\EE\big[ f_{j_1} (\bx_1)\overline{f_{l_2}}(\by_2) \big]
\EE\big[ f_{j_2}(\bx_2)\overline{f_{l_1}}(\by_1) \big]
,
\end{eqnarray}
where the covariance function of the source is given by (\ref{eq:covsource}).

We parameterize  the four points 
$\bx_1,\bx_2,\by_1,\by_2$ in (\ref{def:generalmoment}) in the special way:
\begin{eqnarray}
\label{eq:reliexr1}
\bx_1 = \frac{\br_1+\br_2+\bq_1+\bq_2}{2}, \quad \quad 
\by_1 = \frac{\br_1+\br_2-\bq_1-\bq_2}{2}, \\
\bx_2 = \frac{\br_1-\br_2+\bq_1-\bq_2}{2}, \quad \quad 
\by_2 = \frac{\br_1-\br_2-\bq_1+\bq_2}{2}.
\label{eq:reliexr2}
\end{eqnarray}
In particular $\br_1/2$ is the barycenter of the four points $\bx_1,\bx_2,\by_1,\by_2$:
\begin{eqnarray*}
\br_1 = \frac{\bx_1+\bx_2+\by_1+\by_2}{2} , \quad \quad 
\bq_1 = \frac{\bx_1+\bx_2-\by_1-\by_2}{2}, \\
\br_2 = \frac{\bx_1-\bx_2+\by_1-\by_2}{2}, \quad \quad 
\bq_2 = \frac{\bx_1-\bx_2-\by_1+\by_2}{2}.
\end{eqnarray*}
We denote by $\mu$ the fourth-order moment in these new variables (without writing the dependence on $j_1,j_2,l_1,l_2$):
\begin{equation}
\mu(z,\bq_1,\bq_2,\br_1,\br_2) = 
\mu_{4,j_1l_1j_2l_2}  (z, 
\bx_1  ,
\bx_2 ,
\by_1 ,  
\by_2 
)
\end{equation}
with $\bx_1,\bx_2,\by_1,\by_2$ given by (\ref{eq:reliexr1}-\ref{eq:reliexr2}) in terms of $\bq_1,\bq_2,\br_1,\br_2$.

In the variables $(\bq_1,\bq_2,\br_1,\br_2)$ the function ${\mu} $ satisfies the system:
\begin{equation}
\label{eq:M20}
\frac{\partial {\mu}}{\partial z} = \frac{i}{k_o} \big( \nabla_{\br_1}\cdot \nabla_{\bq_1}
+
 \nabla_{\br_2}\cdot \nabla_{\bq_2}
\big)  {\mu} + \frac{k_o^2}{4} {U}(\bq_1,\bq_2,\br_1,\br_2) {\mu}   ,
\end{equation}
with the generalized potential
\begin{eqnarray}
\nonumber
{U}(\bq_1,\bq_2,\br_1,\br_2) &=& 
{C}(\bq_2+\bq_1) 
+
{C}(\bq_2-\bq_1) 
+
 {C}(\br_2+\bq_1) 
+ 
{C}(\br_2-\bq_1) \\
&&- 
 {C}( \bq_2+\br_2) -  {C}( \bq_2-\br_2) - 2 {C}({\bf 0})  .
\end{eqnarray}
Note in particular that the generalized potential does not depend on the barycenter $\br_1$, and this comes from the fact that the medium is statistically homogeneous.
The Fourier transform (in $\bq_1$, $\bq_2$, $\br_1$, and $\br_2$) of the fourth-order moment
is defined by:
\begin{eqnarray}
\nonumber
\hat{\mu}(z,\bxi_1,\bxi_2,\bzeta_1,\bzeta_2) 
&=& 
\iint_{\RR^2\times \RR^2} {\mu}(z,\bq_1,\bq_2,\br_1,\br_2)  \\
&& \hspace*{-0.8in}
\times
\exp  \big(- i\bq_1 \cdot \bxi_1- i\bq_2 \cdot \bxi_2- i\br_1\cdot \bzeta_1- i\br_2\cdot \bzeta_2\big) d\bq_1d\bq_2 
d\br_1d\br_2 \label{eq:fourier} 
. \hspace*{0.3in} 
\end{eqnarray}
It satisfies
\begin{eqnarray}
\nonumber
&& 
\frac{\partial \hat{\mu}}{\partial z} + \frac{i}{k_o} \big( \bxi_1\cdot \bzeta_1+   \bxi_2\cdot \bzeta_2\big) \hat{\mu}
=
\frac{k_o^2}{4 (2\pi)^2} 
\int_{\RR^2}  \hat{C}(\bk) \bigg[  
 \hat{\mu} (  \bxi_1-\bk, \bxi_2-\bk, \bzeta_1, \bzeta_2)  \\
\nonumber
&& \quad   + \hat{\mu} (  \bxi_1+\bk, \bxi_2-\bk, \bzeta_1, \bzeta_2)
  -
2 \hat{\mu}(\bxi_1,\bxi_2, \bzeta_1, \bzeta_2)    \\
\nonumber
&&  \quad 
+   \hat{\mu} (  \bxi_1+\bk,\bxi_2, \bzeta_1,  \bzeta_2-\bk)  + 
  \hat{\mu} (  \bxi_1-\bk,\bxi_2,  \bzeta_1, \bzeta_2-\bk)    
\\
&&  \quad 
-  \hat{\mu} (  \bxi_1,\bxi_2-\bk, \bzeta_1, \bzeta_2-\bk)  
-  \hat{\mu} (  \bxi_1,\bxi_2+\bk,  \bzeta_1, \bzeta_2-\bk) 
\bigg]d \bk .
\label{eq:fouriermom0}
\end{eqnarray}
If $j_1=j_2=l_2=l_2 \equiv j \in \{1,2\}$,
then the initial condition is 
\begin{eqnarray}
\nonumber
\hat{\mu}(z=0,\bxi_1,\bxi_2,\bzeta_1,\bzeta_2) &=& 
(2\pi)^4 A_j^4 \phi_{\rho_o}^1(\bxi_1)  \phi_{\rho_o}^1(\bxi_2)  \phi_{r_o}^1(\bzeta_1)  \phi_{r_o}^1(\bzeta_2)
\\
&& +
(2\pi)^4  A_j^4  \phi_{\rho_o}^1(\bxi_1)  \phi_{\rho_o}^1(\bzeta_2)  \phi_{r_o}^1(\bzeta_1)  \phi_{r_o}^1(\bxi_2)  ,
\end{eqnarray}
with
\begin{equation}
\phi^1_{\rho}(\bxi) = \frac{\rho^2}{2\pi } \exp \Big( - \frac{\rho^2 |\bxi|^2}{2}\Big)  .
\end{equation}
Similar Gaussian expressions hold for the initial condition in the other cases for $(j_1,j_2,l_1,l_2)$, we only address the case $j_1=j_2=l_2=l_2 \equiv j $ in the following.
%We look for the second-order moment of the intensity that is
%\begin{eqnarray}\label{eq:mu4b} 
%\mu_4(z,\bx_1,\bx_2,\bx_1,\bx_2 )&=& \mu(z,{\bf 0},{\bf 0},\br_1=\bx_1+\bx_2,\br_2=\bx_1-\bx_2 )\\
%&=& \frac{1}{(2\pi)^8}
%\iint \hat{\mu}(z,\bxi_1,\bxi_2,\bzeta_1,\bzeta_2) e^{i \br_1\cdot\bzeta_1+i\br_2\cdot\bzeta_2} d\bxi_1 d\bxi_2 d\bzeta_1 d\bzeta_2 .
%\nonumber
%\end{eqnarray}

We cannot solve the problem for the  fourth moment $\mu$ explicitly and consider next a secondary
scaling  limit where we can identify  an explicit solution. 
We consider the scintillation regime, 
discussed in more detail in Appendix \ref{app:a},
where  the correlation radius of the source is of the same order as 
the correlation radius of the medium, but the beam radius of the source is much larger:
\begin{equation}
\rho_o \to  {\rho_o} ,\quad C(\bx) \to \eps C(\bx), \quad r_o\to \frac{r_o}{\eps}, \quad z \to \frac{z}{\eps}.
\end{equation}
We introduce the rescaled function
\begin{equation}
\tilde{\mu}^\eps (z , \bxi_1,\bxi_2 ,\bzeta_1,\bzeta_2 ) = \mu \Big(\frac{z}{\eps} , \bxi_1,\bxi_2 ,\bzeta_1,\bzeta_2  \Big)
\exp \Big( i \frac{z}{k_o\eps} (\bxi_1\cdot\bzeta_1+\bxi_2\cdot \bzeta_2) \Big).
\end{equation}
 Then the limit  $\eps\to 0$ is applied, corresponding to the scintillation regime.

In the scintillation regime the rescaled function $\tilde{\mu}^\eps$ 
satisfies the equation with fast phases
\begin{eqnarray}
\nonumber
 \frac{\partial \tilde{\mu}^\eps}{\partial z} 
&=&
\frac{k_o^2}{4 (2\pi)^2} 
\int_{\RR^2}  \hat{C}(\bk) \bigg[  - 2 \tilde{\mu}^\eps(  \bxi_1 ,\bxi_2, \bzeta_1,\bzeta_2)   \\
\nonumber
&&+
\tilde{\mu}^\eps (  \bxi_1-\bk,\bxi_2-\bk,  \bzeta_1,\bzeta_2) 
e^{i\frac{c_o z}{\eps \omega_o} \bk \cdot  (\bzeta_2 + \bzeta_1)} \\
&& \nonumber
+ 
\tilde{\mu}^\eps  (  \bxi_1-\bk,\bxi_2,  \bzeta_1,\bzeta_2-\bk) 
e^{i\frac{c_o z}{\eps \omega_o} \bk \cdot ( \bxi_2 +  \bzeta_1)}\\
\nonumber
&& +\tilde{\mu}^\eps (  \bxi_1+\bk,\bxi_2-\bk,  \bzeta_1,\bzeta_2) 
e^{i\frac{c_o z}{\eps \omega_o} \bk \cdot (\bzeta_2 -   \bzeta_1)}\\
&& \nonumber
+ 
\tilde{\mu}^\eps  (  \bxi_1+\bk,\bxi_2,  \bzeta_1,\bzeta_2-\bk) 
e^{i\frac{c_o z}{\eps \omega_o} \bk \cdot (\bxi_2 -  \bzeta_1)}\\
\nonumber
&&
-
 \tilde{\mu}^\eps  (  \bxi_1,\bxi_2-\bk, \bzeta_1, \bzeta_2-\bk) 
 e^{i\frac{c_o z}{\eps \omega_o} (  \bk \cdot (\bzeta_2+\bxi_2)-|\bk|^2 )} \\
 &&
- \tilde{\mu}^\eps  ( \bxi_1,\bxi_2-\bk,  \bzeta_1,\bzeta_2+\bk) 
e^{i\frac{c_o z}{\eps \omega_o} ( \bk \cdot (\bzeta_2-\bxi_2)+|\bk|^2)}
\bigg] d \bk ,
\label{eq:tildeNeps}
\end{eqnarray}
starting from 
\begin{eqnarray}
\nonumber
\tilde{\mu}^\eps (z=0,\bxi_1,\bxi_2, \bzeta_1, \bzeta_2 ) &=& (2\pi)^8 A_j^4 \phi^1_{\rho_o} ( \bxi_1 )
\phi^1_{\rho_o} ( \bxi_2 )
\phi^\eps_{r_o} ( \bzeta_1 )\phi^\eps_{r_o} ( \bzeta_2 )\\ &&
+ (2\pi)^8A_j^4  \phi^1_{\rho_o} ( \bxi_1 )
\phi^1_{\rho_o} ( \bzeta_2 )
\phi^\eps_{r_o} ( \bzeta_1 )\phi^\eps_{r_o} ( \bxi_2 ) ,
\label{eq:initialtildeM2eps}
\end{eqnarray}
where $\phi^\eps_\rho$ is defined by:
\begin{equation}
\label{def:phiepsrho}
\phi^\eps_{\rho}(\bxi) = \frac{\rho^2}{2\pi \eps^2} \exp \Big( -\frac{\rho^2}{2 \eps^2} |\bxi|^2\Big) .
\end{equation}
The following result  shows that $\tilde{\mu}^\eps$ exhibits a multi-scale behavior
as $\eps \to 0$, with some components evolving at the scale $\eps$ and 
some components evolving on the order one scale \cite{gar16a}.
\begin{proposition}
\label{prop:3}%
The function $\tilde{\mu}^\eps (z,\bxi_1,\bxi_2, \bzeta_1,\bzeta_2 ) $ can be expanded as
\begin{eqnarray}
\nonumber
 \tilde{\mu}^\eps\big( z ,\bxi_1,\bxi_2,  \bzeta_1,\bzeta_2 \big)  &=&
 (2\pi)^8 A_j^4 
\phi^\eps_{r_o} ( \bzeta_1 )
\phi^\eps_{r_o} ( \bzeta_2 )
B\big(z, \bxi_1,\bxi_2 ,\frac{\bzeta_1}{\eps},\frac{\bzeta_2}{\eps} \big) \\
\nonumber
&& +
 (2\pi)^8A_j^4  \phi^\eps_{r_o} ( \bzeta_1 )
\phi^\eps_{r_o} ( \bxi_2 )
B\big(z, \bxi_1,\bzeta_2 ,\frac{\bzeta_1}{\eps},\frac{\bxi_2}{\eps} \big) \\
&& 
 + R^\eps (z ,\bxi_1,\bxi_2 ,  \bzeta_1 ,\bzeta_2 )   ,
\label{eq:propsci11}
\end{eqnarray}
where 
\begin{eqnarray}
\nonumber
&&
B(z,\bxi_1,\bxi_2,\bzeta_1,\bzeta_2)
=\frac{1}{(2\pi)^4}
\iint_{\RR^2\times \RR^2} d\bx d\by\exp\Big( - \frac{|\bx|^2+|\by|^2}{2\rho_o^2}-i \bxi_1\cdot\bx -i \bxi_2\cdot \by \\
&& +\frac{k_o^2}{4}
\int_0^z C\big(\bx+\by+\frac{z'}{k_o}(\bzeta_1+\bzeta_2)\big)
+C\big(\bx-\by+\frac{z'}{k_o}(\bzeta_1-\bzeta_2)\big)
-2C({\bf 0}) dz' \Big) ,
\end{eqnarray}
and the function $R^\eps$ goes to  $0$ as $\eps\to0$. 
%satisfies
%\begin{equation}
%\sup_{z \in [0,Z]} \| R^\eps (z,\cdot,\cdot,\cdot,\cdot) \|_{L^1(\RR^2\times \RR^2\times \RR^2\times \RR^2)} 
%\stackrel{\eps \to 0}{\longrightarrow}  0  ,
%\end{equation}
%for any $Z>0$.
\end{proposition}

As a consequence, the second-order moment of the intensity is 
\begin{eqnarray}
 \EE \big[ \big|u\big( \frac{z}{\eps}, \frac{\br}{\eps}+\frac{\bq}{2}\big)\big|^2 \big|u\big(\frac{z}{\eps},\frac{\br}{\eps}-\frac{\bq}{2}\big) \big|^2  \big] =A_j^4 
 {\cal I}_{\rho_o}(z,2\br,{\bf 0})+A_j^4 {\cal J}_{\rho_o}(z,2\br,\bq),
 \end{eqnarray}
 with
 \begin{eqnarray}
\nonumber
&&{\cal I}_{\rho_o}(z,\br_1,\br_2) 
\\
\nonumber 
&&=  \iint_{\RR^2 \times \RR^2} d\bxi_1d\bxi_2d\bzeta_1d\bzeta_2\phi_{r_o}^1(\bzeta_1)\phi_{r_o}^1(\bzeta_2) B(z,\bxi_1,\bxi_2,\bzeta_1,\bzeta_2)\\
\nonumber
&& \quad \times \exp\Big( -i \frac{z}{k_o} (\bxi_1\cdot\bzeta_1+\bxi_2\cdot\bzeta_2) +  i \bzeta_1 \cdot \br_1+  i \bzeta_2 \cdot \br_2 \Big)   \\
\nonumber&&=
\Big( \int_{\RR^2}  \phi^1_{\frac{r_o}{\sqrt{2}}}(\bzeta) \exp \Big( i  \bzeta \cdot \frac{\br_1 +\br_2}{2}+\frac{k_o^2}{4}\int_0^z C(\frac{\bzeta z'}{k_o})-C({\bf 0}) dz'
-\frac{|\bzeta|^2 z^2}{4 k_o^2 \rho_o^2} \Big) d\bzeta\Big) \\
\nonumber
&& \quad \times 
\Big( \int_{\RR^2}  \phi^1_{\frac{r_o}{\sqrt{2}}}(\bzeta) \exp \Big( i  \bzeta \cdot \frac{\br_1 - \br_2}{2} +\frac{k_o^2}{4}\int_0^z C(\frac{\bzeta z'}{k_o})-C({\bf 0}) dz'
-\frac{ |\bzeta|^2 z^2 }{4k_o^2 \rho_o^2} \Big) d\bzeta\Big) \\
&& = {\cal H}_{\rho_o}(z,\frac{\br_1+\br_2}{2},{\bf 0}){\cal H}_{\rho_o}(z,\frac{\br_1-\br_2}{2},{\bf 0})
\label{eq:defIrho0}
\end{eqnarray}
and
\begin{eqnarray}
\nonumber
&&
{\cal J}_{\rho_o}(z,\br_1,\br_2) \\
\nonumber
&&= 
 \iint_{\RR^2 \times \RR^2}d\bxi_1d\bxi_2d\bzeta_1d\bzeta_2\phi_{r_o}^1(\bzeta_1)\phi_{r_o}^1(\bxi_2) B(z,\bxi_1,\bzeta_2,\bzeta_1,\bxi_2)\\
\nonumber
&&\quad \times \exp\Big( -i \frac{z}{k_o} (\bxi_1\cdot\bzeta_1+\bxi_2\cdot\bzeta_2) +  i \bzeta_1 \cdot \br_1 +i\bzeta_2 \cdot \br_2  \Big)   \\
\nonumber
&&=
\Big| \int_{\RR^2}  \phi^1_{\frac{r_o}{\sqrt{2}}}(\bzeta) \exp \Big( i \bzeta \cdot \frac{\br_1}{2}  +\frac{k_o^2}{4}\int_0^z C(\frac{\bzeta z'}{k_o}-\br_2)-C({\bf 0}) dz'
-\frac{| \frac{\bzeta z}{k_o}-\br_2|^2}{4\rho_o^2} \Big) d\bzeta\Big|^2 \\
&&= \big| {\cal H}_{\rho_o}(z,\frac{\br_1}{2},\br_2) \big|^2 ,
\label{eq:defJrho0} 
\end{eqnarray}
where ${\cal H}_{\rho_o}$ is defined by (\ref{def:Hq}). 
We finally remark that for far away points the second-order moment of the intensity is 
\begin{eqnarray*}
&&  \EE \left[ \big|u\big( \frac{z}{\eps}, \frac{\br}{\eps}+\frac{\bq}{2\eps}\big)\big|^2 \big|u\big(\frac{z}{\eps},\frac{\br}{\eps}-\frac{\bq}{2\eps}\big) \big|^2  \right] 
=
\big(A_1^2+A_2^2\big)^2  {\cal I}_{\rho_o} (z,2\br,\bq)    
   \\  && ~~~~~~~~~
    \hbox{}  = \big(A_1^2+A_2^2\big)^2 {\cal H}_{\rho_o}(z, \br +\frac{\bq}{2},{\bf 0})  {\cal H}_{\rho_o}(z, \br -\frac{\bq}{2},{\bf 0}) 
       \\  && ~~~~~~~~~
  =  \EE \left[ \big|u\big( \frac{z}{\eps}, \frac{\br}{\eps}+\frac{\bq}{2\eps}\big)\big|^2  \right]
  \EE \left[ 
  \big|u\big(\frac{z}{\eps},\frac{\br}{\eps}-\frac{\bq}{2\eps}\big) \big|^2  \right] ,
\end{eqnarray*}
so that the intensities then indeed are uncorrelated.

\section*{Acknowledgements}

JG was supported by the Agence Nationale pour la Recherche under Grant No. ANR-19-CE46-0007 (project ICCI), 
 and Air Force Office of Scientific Research under grant FA9550-18-1-0217.
 \\
KS was supported by the Air Force Office of Scientific Research under grant FA9550-18-1-0217,  and   the National Science Foundation under grant DMS-2010046.

\end{document}